# Antipodal Interval-Valued Fuzzy Graphs

**Hossein Rashmanlou[1] and Madhumangal Pal [2]**

[1]Department of Mathematics, Islamic Azad University, Central Tehran Branch, Tehran, Iran
Email:hrashmanlou@yahoo.com

[2]Department of Applied Mathematics with Oceanology and Computer Programming, Vidyasagar University, Midnapore-721102, India
Email: mmpalvu@gmail.com

**Abstract**

***Concepts of graph theory have applications in many areas of computer science including data mining, image segmentation, clustering, image capturing, networks, etc . An interval-valued fuzzy set is a generalization of the notion of a fuzzy set. Interval-valued fuzzy models give more precision, flexibility and compatibility to the system as compared to the fuzzy models. In this paper, we introduce the concept of antipodal interval - valued fuzzy graph and self median interval-valued fuzzy graph of the given interval-valued fuzzy graph. We investigate isomorphism properties of antipodal interval - valued fuzzy graphs.***

## 1 Introduction

The major role of graph theory in computer applications is the development of graph algorithms. A number of algorithms are used to solve problems that are modeled in the form of graphs. These algorithms are used to solve the graph theoretical concepts, which in turn are used to solve the corresponding computer science application problems. Several computer programming languages support the graph theory concepts .The main goal of such languages is to enable the user to formulate operations on graphs in a compact and natural manner. Some of these languages are (1) SPANTREE: to find a spanning tree in the given graph, (2) GTPL: graph theoretic language, (3) GASP: graph algorithm software package, (4) HINT: an extension of LISP. In 1975, Zadeh [31] introduced the notion of interval- valued fuzzy sets as an extension of fuzzy sets [32] in which the values of the membership degrees are intervals of numbers instead of the numbers. Interval- valued fuzzy sets provide a more adequate description of uncertainty than traditional fuzzy sets. It is therefore important to use interval-valued fuzzy sets in applications, such as fuzzy control. One of the computationally most intensive parts of fuzzy control is defuzzification [8].

The fuzzy graph theory as a generalization of Euler's graph theory was first introduced by Rosenfeld [12] in 1975. The fuzzy relations between fuzzy sets were first considered by Rosenfeld, who developed the structure of fuzzy graphs obtaining analogues to several graph theoretical concepts. Later, Bhattacharya [7] gave some remarks on fuzzy graphs, and some operations on fuzzy graphs were introduced by Modeson and Peng [9]. The complement of a fuzzy graph was defined by Mordeson [10] and future studied by Sunitha and Kumar [20].

Recently, Akram et al. introduced the concepts of bipolar fuzzy graphs, interval - valued fuzzy graphs, strong intuitionistic fuzzy graphs in [1-5].

Talebi and Rashmanlou [25] studied properties of isomorphism and complement on interval-valued fuzzy graphs. Likewise, they defined isomorphism and some new operations on vague graphs [26-27]. Rashmanlou and Jun defined complete interval - valued fuzzy graphs [13]. Talebi, Rashmanlou and Davvaz in [28] investigated some properties of interval- valued fuzzy graphs such as regular interval- valued fuzzy graph, totally regular interval-valued fuzzy graph and complement of interval-valued fuzzy graph. Talebi and Rashmanlou defined product bipolar fuzzy graphs [29] and isomorphism and complement on bipolar fuzzy graphs [30].

Recently Rashmanlou and Pal defined irregular interval - valued fuzzy graphs [11]. More results on interval - valued fuzzy graphs [14], product interval- valued fuzzy graphs and their degrees [15], intuitionistic fuzzy graphs with categorical properties [17], some properties of highly irregular interval - valued fuzzy graphs [18], A study on bipolar fuzzy graphs [19] and investigated several properties. They defined isometry on interval-valued fuzzy graphs [14].

Samanta and Pal introduced fuzzy tolerance graph [21], irregular bipolar fuzzy graphs [23], fuzzy k- competition graphs and *P*-competition fuzzy graphs [24], bipolar fuzzy hypergraphs [22].

Since interval-valued fuzzy set theory is an increasingly popular extension of fuzzy set theory where traditional [0, 1]-valued membership degrees are replaced by intervals in [0, 1] that approximate the (unknown) membership degrees; specific types of interval valued fuzzy graphs have been introduced and investigated. In this paper we introduce the concept of an antipodal interval-valued fuzzy graph and self median interval- valued fuzzy graph of the given interval-valued fuzzy graph. We also investigate isomorphism properties of antipodal interval - valued fuzzy graphs. The natural extension of this research work is the application of interval-valued fuzzy graphs in the area of soft computing including neural networks, expert systems, database theory, and geographical information systems.

## 2. Preliminaries

In this section we recall some basic concepts that are necessary for subsequent discussion.

By a graph, we mean a pair $G^* = (V, E)$, where V is the set and E is a relation on V. The elements of V are vertices of $G^*$ and the elements of E are edges of $G^*$. We write $x, y \in E$ to mean $\{x, y\} \in E$, and if $e = xy \in E$, we say x and y are adjacent. Formally, given a graph $G^* = (V, E)$, two vertices $x, y \in V$ are said to be neighbors, or adjacent nodes, if $(x, y) \in E$.

The antipodal graph of a graph $G^*$, denoted by $A(G^*)$, has the same vertex set as $G^*$ with an edge joining vertices u and v if $d(u, v)$ is equal to the diameter of $G^*$. For a graph $G^*$ of order P, the antipodal graph $A(G^*) = G^*$ if and only if $G^* = K_p$. If $G^*$ is a non-complete graph of order P, then $A(G^*) \subset \overline{G^*}$, for a graph $G^*$, the antipodal graph $A(G^*) = \overline{G^*}$ if and only if (a) $G^*$ is of diameter 2 or (b) $G^*$ is disconnected and the components of $G^*$ are complete graphs. A graph $G^*$ is an antipodal graph if and only if it is the antipodal graph of its complement. The self median fuzzy graph were introduced by Ahmed and Gani in [6]. The median of a graph $G^*$ is the set of all vertices v of $G^*$ for which the value $d_{G^*}(v)$ is minimized. A graph $G^*$ is self-median if and only if the value $d_{G^*}(v)$ is constant over all vertices v of $G^*$. The status, or distance sum, of a given vertex v in a graph is defined by $S(v) = \sum_{v \neq u} d(u,v)$, where $d(u,v)$ is the distance from a vertex u to v. In other words, a self median graph $G^*$ is one in which all the nodes have the same status S(v). The graphs $C_n, K_{n,n}$ and $K_n$ are self median. The status of a vertex $v_i$ is denoted by $S(v_i)$ and is defined as $S(v_i) = \sum_{v_i \in V} \delta(v_i, v_j)$. The total status of a fuzzy graph $G^*$ is denoted by $t[S(G^*)]$ and is defined as $t[S(G^*)] = \sum_{v_i \in V} S(v_i)$. The median of a fuzzy graph $G^*$, denoted, is the set of nodes with minimum status. A fuzzy graph $G^*$ is said to be self - median if all the vertices have the same status. By a fuzzy subset $\mu$ on a set X is mean a map $\mu : X \to [0,1]$. A map $\upsilon : X \times X \to [0,1]$ is called a fuzzy relation on X if $\upsilon(x,y) \leq \min(\mu(x), \mu(y))$ for all $x, y \in X$. A fuzzy relation $\upsilon$ is symmetric if $\upsilon(x,y) = \upsilon(y,x)$ for all $x, y \in X$.

**Definition 2.1:** The interval - valued fuzzy set A in V is defined by

$$A = \{ (x, [\mu_A(x), \upsilon_A(x)]) \mid x \in V \},$$

where $\mu_A(x)$ and $\upsilon_A(x)$ are fuzzy subsets of V such that $0 \le \mu_A(x) \le \upsilon_A(x) \le 1$ for all $x \in V$. If $G^* = (V, E)$ is a graph, then by an interval - valued fuzzy relation B on a set E we mean an interval - valued fuzzy set such that

$$\mu_B(xy) \le \min\left(\mu_A(x), \mu_A(y)\right),$$
$$\upsilon_B(xy) \le \min\left(\upsilon_A(x), \upsilon_A(y)\right)$$

for all $xy \in E$.

**Definition 2.2:** By an interval - valued fuzzy graph of a graph $G^* = (V, E)$ we mean a pair $G = (A, B)$, where $A = [\mu_A, \upsilon_A]$ is an interval-valued fuzzy set on V and $B = [\mu_B, \upsilon_B]$ is an interval - valued fuzzy relation on E such that

$$\mu_B(xy) \le \min\left(\mu_A(x), \mu_A(y)\right),$$
$$\upsilon_B(xy) \le \min\left(\upsilon_A(x), \upsilon_A(y)\right).$$

**Definition 2.3:** The complement of an interval - valued fuzzy graph $G = (A, B)$ is an interval - valued fuzzy graph, where

$(i)$ $\overline{V} = V$,

$(ii)$ $\overline{\mu_A}(\upsilon_i) = \mu_A(\upsilon_i)$ and $\overline{\upsilon_A}(\upsilon_i) = \upsilon_A(\upsilon_i)$ for all $v_i \in V$,

$(iii)$ $\overline{\mu_B}(v_i v_j) = \mu_A(v_i) \wedge \mu_A(v_j) - \mu_B(v_i v_j)$, $\overline{\upsilon_B}(v_i v_j) = \upsilon_A(v_i) \wedge \upsilon_A(v_j) - \upsilon_B(v_i v_i)$, for all $v_i, v_j \in V$.

**Definition 2.4:** An interval - valued fuzzy graph G is called complete if $\mu_B(xy) = \min\left(\mu_A(x), \mu_A(y)\right)$ and $\upsilon_B(xy) = \min\left(\upsilon_A(x), \upsilon_A(y)\right)$, for each edge $xy \in E$.

**Definition 2.5:** A path P in an interval - valued fuzzy graph G is a sequence of distinct vertices $v_1, v_2, \ldots, v_n$ such that either one of the following conditions is satisfied:

(i) $\mu_B(xy) > 0$ and $\mu_B(xy) = 0$ for some x , y.

(ii) $\upsilon_B(xy) > 0$ and $\upsilon_B(xy) > 0$ for some x,y. A path P = $v_1, v_2, \ldots, v_{n+1}$ in G is called a cycle if $v_1 = v_{n+1}$ and $n \ge 3$.

**Definition 2.6:** Let P = $v_0, v_1, v_2, ..., v_n$ be a path in interval- valued fuzzy graph G. The $\mu$- strength of the paths connecting any two vertices $v_i, v_j$ is defined as $\max(\mu_B(v_i, v_j))$ and is denoted by $(\mu_B(v_i v_j))^\infty$. The $\upsilon$- strength of the paths connecting any two vertices $v_i, v_j$ is defined as $\max(\upsilon_B(v_i, v_j))$ and is denoted by $(\upsilon_B(v_i, v_j))^\infty$. If same edge possesses both the $\mu$- strength and $\upsilon$- strength value, then it is the strength of the strongest path P and is denoted by $S_p = [(\mu_B(v_i v_j))^\infty, (\upsilon_B(v_i v_j))^\infty]$ for all i,j= 1,2,…, n.

**Definition 2.7:** An interval - valued fuzzy graph G is connected if any two vertices are joined by a path. That is, an interval - valued fuzzy graph G is connected if $(\mu_B(v_i v_j))^\infty > 0$ and $(\upsilon_B(v_i v_j))^\infty > 0$.

**Definition 2.8:** Let G be a connected interval - valued fuzzy graph. The $\mu$ - length of a path P: $v_1, v_2, ..., v_n$ in G, $L_\mu(p)$, is defined as $L_\mu(p) = \sum_{i=1}^{n-1} \mu_B(v_i, v_{i+1})$.

The $\upsilon$- length of a path P: $v_1, v_2, ..., v_n$ in G , $L_\upsilon(p)$, is defined as

$L_\upsilon(p) = \sum_{i=1}^{n-1} \upsilon_B(v_i, v_{i+1})$. The $\mu\upsilon$ - length of a path P: $v_1, v_2, …, v_n$ in G, $L_{\mu\upsilon}(p)$, is defined as $L_{\mu\upsilon}(p) = [L_\mu, L_\upsilon]$.

**Definition 2.9:** Let G be a connected interval - valued fuzzy graph. The $\mu$- distance, $\delta_\mu(v_i, v_j)$, is the smallest $\mu$- length of any $v_i$ - $v_j$ path P in G, where $v_i, v_j \in V$. That is, $\delta_\mu(v_i, v_j) = \min(L_\mu(P))$. The $\upsilon$ - distance, $\delta_\upsilon(v_i, v_j)$, is the largest $\upsilon$ - length of any $v_i$-$v_j$ path P in G, where $v_i, v_j \in V$. That is, $\delta_\upsilon(v_i, v_j) = \max(L_\upsilon(P))$. The distance, $\delta(v_i, v_j)$, is defined as $\delta(v_i, v_j) = [\delta_\mu(v_i, v_j), \delta_\upsilon(v_i, v_j)]$.

**Definition 2.10:** Let G be a connected interval - valued fuzzy graph for each $v_i \in V$, the $\mu$ - eccentricity of $v_i$, denoted by $e_\mu(v_i)$, is defined as $e_\mu(v_i) = \max\left\{\delta_\mu(v_i, v_j) \mid v_i \in V, v_i \neq v_j\right\}$. For each $v_i \in V$, the $\upsilon$- eccentricity of $v_i$, denoted by $e_\upsilon(v_i)$, is defined as $e_\upsilon(v_i) = \max\left\{\delta_\upsilon(v_i, v_j) \mid v_i \in V, v_i \neq v_j\right\}$. For each $v_i \in V$, the eccentricity of $v_i$, denoted by $e(v_i)$, is defined as $e(v_i) = [e_\mu(v_i), e_\upsilon(v_j)]$.

**Definition 2.11:** Let G be a connected interval - valued fuzzy graph. The $\mu$-radius of G is denoted by $r_\mu(G)$ and is defined as $r_\mu(G) = \min\{e_\mu(v_i) \mid v_i \in V\}$. The $\upsilon$-radius of G is denoted by $r_\upsilon(G)$ and is defined by $r_\upsilon(G) = \min\{e_\upsilon(v_i) \mid v_i \in V\}$. The radius of G is denoted by $r(G)$ and is defined as $r(G) = [r_\mu(G), r_\upsilon(G)]$.

**Definition 2.12:** Let G be a connected interval - valued fuzzy graph. The $\mu$-diameter of G is denoted by $d_\mu(G)$ and is defined as $d_\mu(G) = \max\{e_\mu(v_i) \mid v_i \in V\}$. The $\upsilon$-diameter of G is denoted by $d_\upsilon(G)$ and is defined as $d_\upsilon(G) = \max\{e_\upsilon(v_i) \mid v_i \in V\}$. The diameter of G is denoted by $d(G)$ and is defined as $d(G) = [d_\mu(G), d_\upsilon(G)]$.

**Example 2.13:** Consider a connected interval - valued fuzzy graph G such that $V = \{u, v, x, w\}, E = \{(w, x), (w, v), (w, u), (x, v), (u, v)\}$.

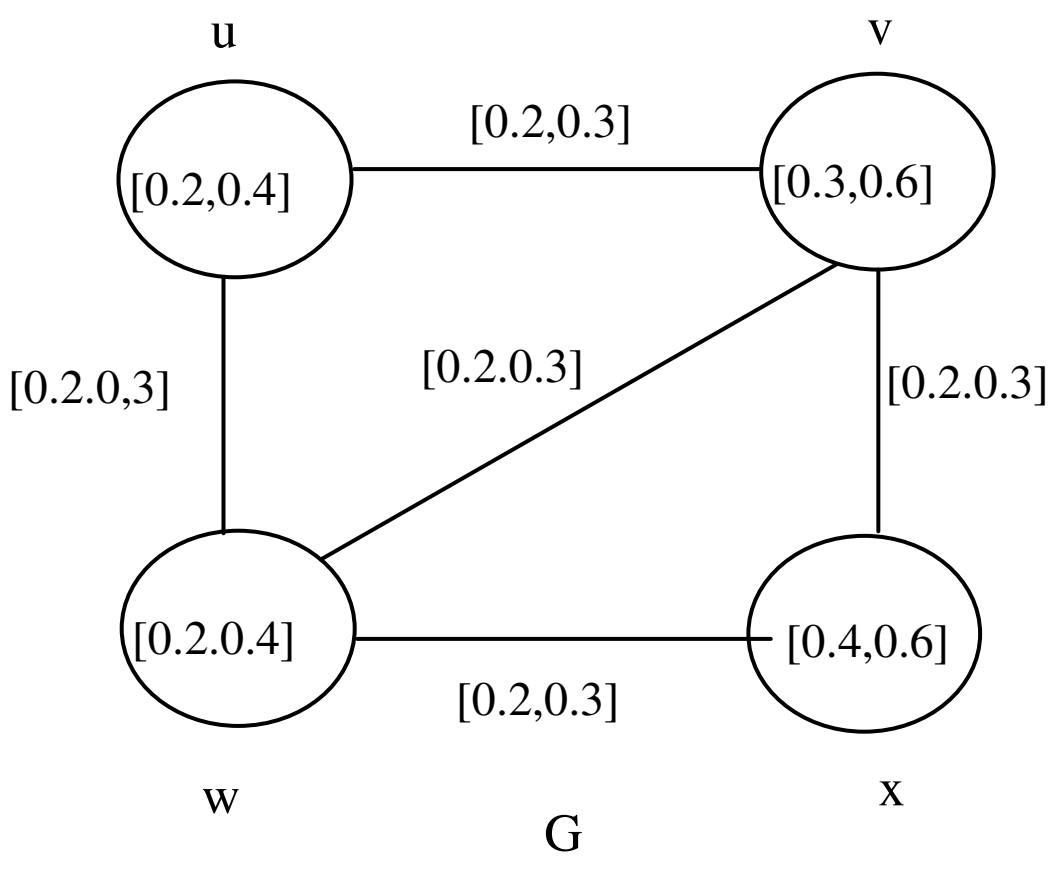

**Fig. 1: Interval - valued fuzzy graph G**

By routine computations, it is easy to see that:

(i)

$\delta_\mu(wu) = 0.2$ , $\delta_\mu(w,x) = 0.2$ , $\delta_\mu(w,v) = 0.2.$

$\delta_\mu(v,x) = 0.2$ , $\delta_\mu(x,u) = 0.4$ , $\delta_\mu(v,u) = 0.2.$

$\delta_\upsilon(w,u) = 0.9$ , $\delta_\upsilon(w,x) = 0.6$ , $\delta_\upsilon(w,v) = 0.6,$

$\delta_\upsilon(v,x) = 0.9$ , $\delta_\upsilon(x,u) = 0.9$ , $\delta_\upsilon(v,u) = 0.9.$

Distance $\delta(v_i, v_j)$ is

$\delta(w,u) = [0.2, 0.9]$ , $\delta(w,x) = [0.2, 0.6]$ , $\delta(w,v) = [0.2, 0.6]$,

$\delta(v,x) = [0.2, 0.9]$ , $\delta(x,u) = [0.4, 0.9]$ , $\delta(v,u) = [0.2, 0.9]$.

(ii) $\mu$-eccentricity and $\upsilon$-eccentricity of the vertices are

$e_\mu(w) = 0.2$ , $e_\mu(x) = 0.4$ , $e_\mu(v) = 0.2$ , $e_\mu(u) = 0.4$,

$e_\upsilon(w) = 0.9$ , $e_\upsilon(x) = 0.9$ , $e_\upsilon(v) = 0.9$ , $e_\upsilon(u) = 0.9.$

The eccentricities of the vertices are

$e(w) = [0.2, 0.9]$ , $e(x) = [0.4, 0.9]$ , $e(v) = [0.2, 0.9]$ , $e(u) = [0.4, 0.9].$

(iii) Radius of G is [0.2, 0.9], diameter of G is [0.4 , 0.9].

## 3. Antipodal interval - valued fuzzy graphs

**Definition 3.1:** Let $G = (A, B)$ be an interval - valued fuzzy graph. An antipodal interval-valued fuzzy graph $A(G) = (E, F)$ is an interval - valued fuzzy graph $G = (A, B)$ in which:

(i) An interval - valued fuzzy vertex set of G is taken as interval - valued fuzzy vertex set of $A(G)$, that is, $\mu_E(x) = \mu_A(x)$ and $\upsilon_E(x) = \upsilon_A(x)$ for all $x \in V$,

(ii) if $\delta(x, y) = d(G)$ , then $\mu_F(xy) = \mu_B(xy)$ if x and y are neighbors in G,

$\mu_F(xy) = \min(\mu_A(x), \mu_A(y))$ if x and y are not neighbors in G, $\upsilon_F(xy) = \upsilon_B(xy)$ if x and y are neighbors in G, $\upsilon_F(xy) = \min(\upsilon_A(x), \upsilon_A(y))$ if x and y are not neighbors in G.

**Example 3.2:** Consider an interval - valued fuzzy graph G such that $A = \{v_1, v_2, v_3, v_4\}$ , $B = \{v_1v_2, v_2v_3, v_3v_4, v_4v_1\}$.

By routine calculations, we have

$\delta_\mu(v_1, v_2) = 0.1$ , $\delta_\mu(v_1, v_3) = 0.2$, $\delta_\mu(v_1, v_4) = 0.1$ ,

$\delta_\mu(v_2, v_3) = 0.1$, $\delta_\mu(v_2, v_4) = 0.2$, $\delta_\mu(v_3, v_4) = 0.1$ ,

$\delta_\upsilon(v_1, v_2) = 0.6$ , $\delta_\upsilon(v_1, v_3) = 0.4$, $\delta_\upsilon(v_1, v_4) = 0.6$,

$\delta_\upsilon(v_2, v_3) = 0.6$, $\delta_\upsilon(v_2, v_4) = 0.4$, $\delta_\upsilon(v_3, v_4) = 0.6$.

$e_\mu(v_1) = 0.2$, $e_\mu(v_2) = 0.2$ , $e_\mu(v_3) = 0.2$, $e_\mu(v_4) = 0.2$,

$e_\upsilon(v_1) = 0.6$, $e_\upsilon(v_2) = 0.6$, $e_\upsilon(v_3) = 0.6$, $e_\upsilon(v_4) = 0.6$.

$d(G) = (0.2, 0.6)$

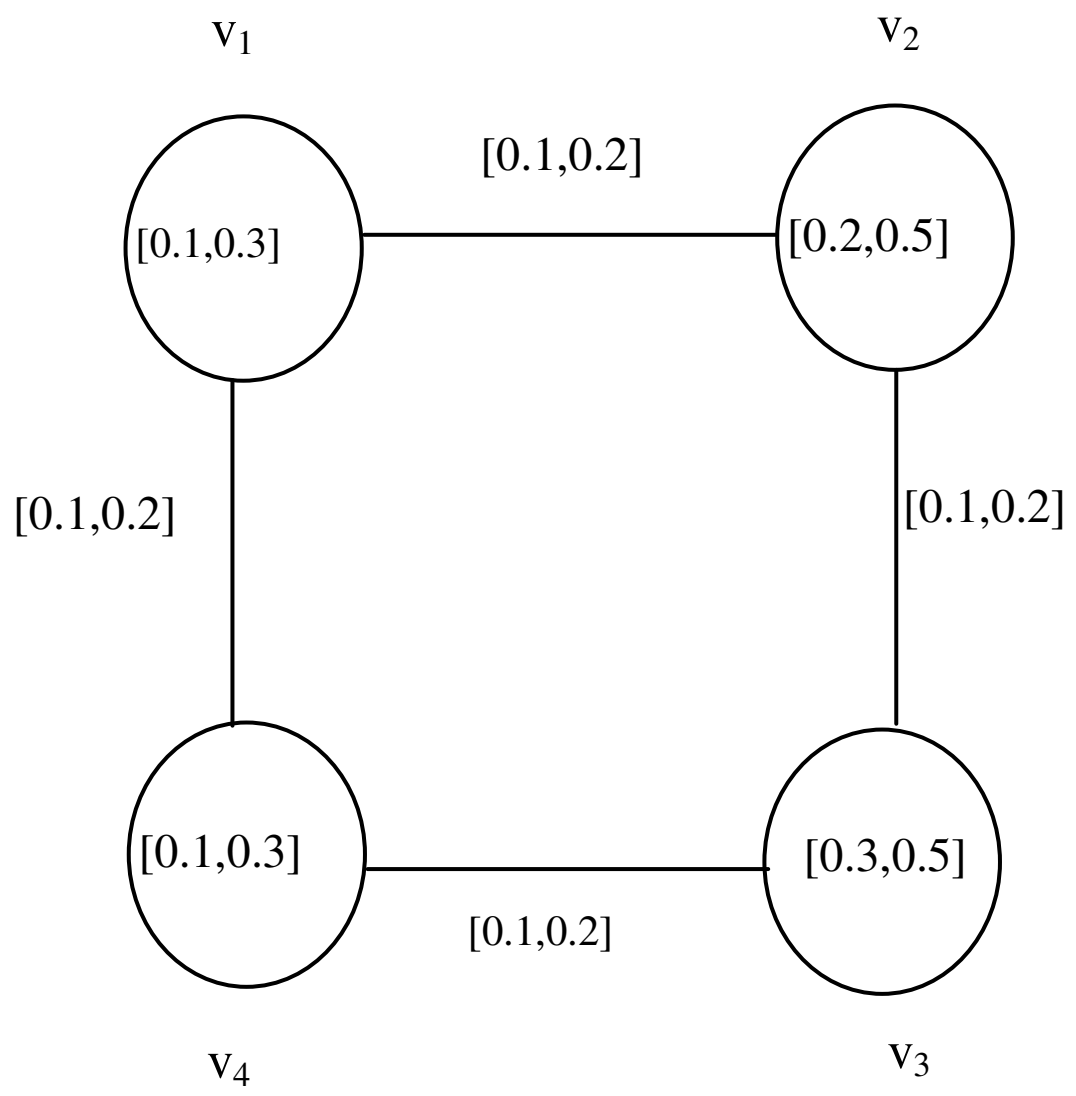

**Fig. 2 (a): Interval - valued fuzzy graph G**

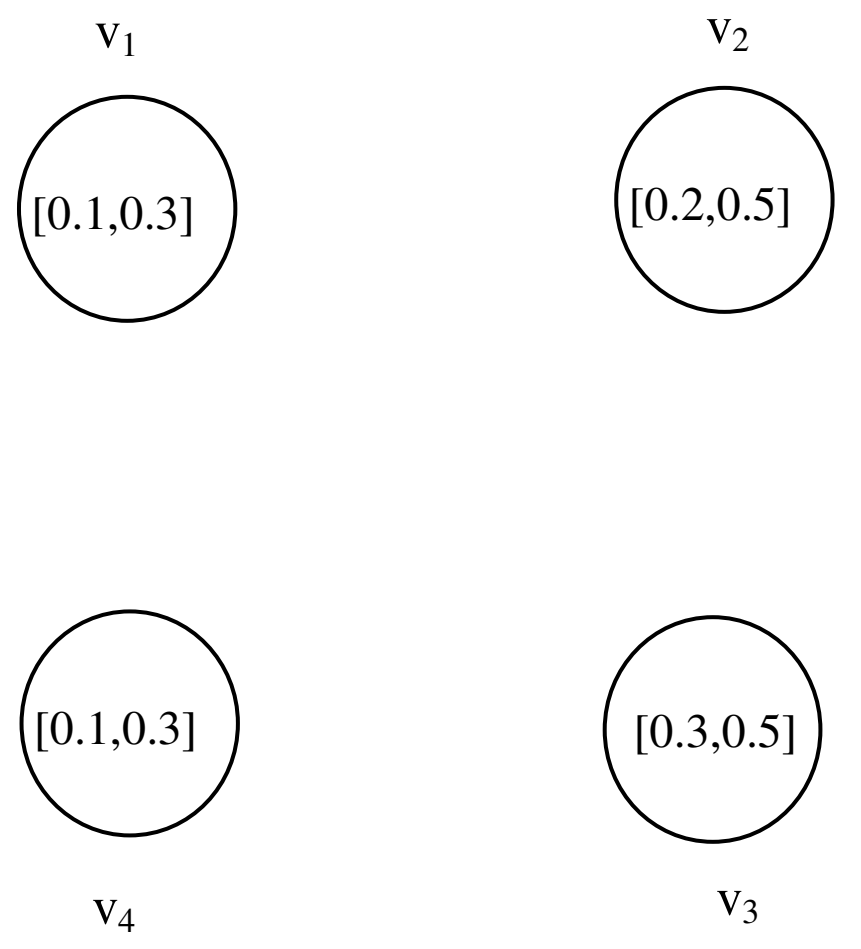

**Fig 2 (b): Antipodal interval - valued fuzzy graph G**

Hence $A(G) = (E, F)$ such that $E = \{v_1, v_2, v_3, v_4\}$ and $F = \phi$.

**Example 3.3:** Consider an interval - valued fuzzy graph G such that $A = \{v_1, v_2, v_3\}$ , $B = \{v_1v_2, v_1v_3, v_2v_3\}$ .

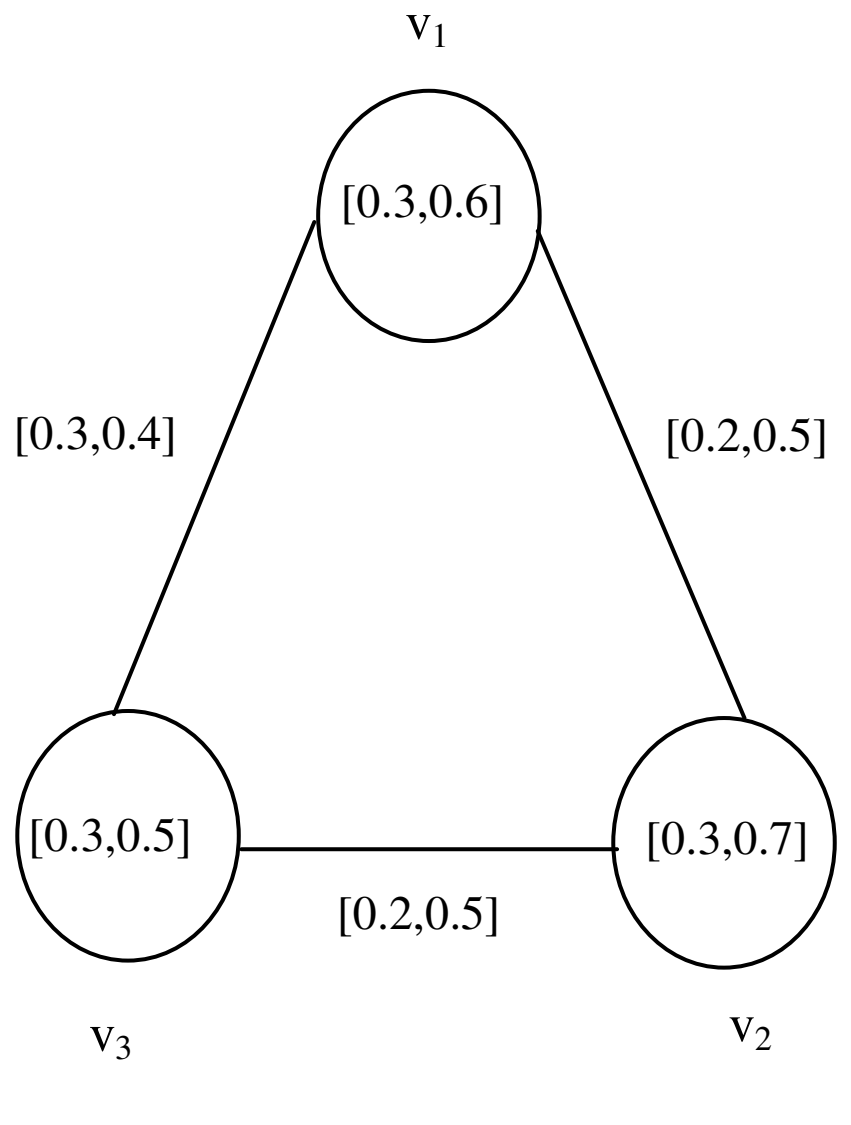

**Fig. 3 (a): Interval - valued fuzzy graph G**

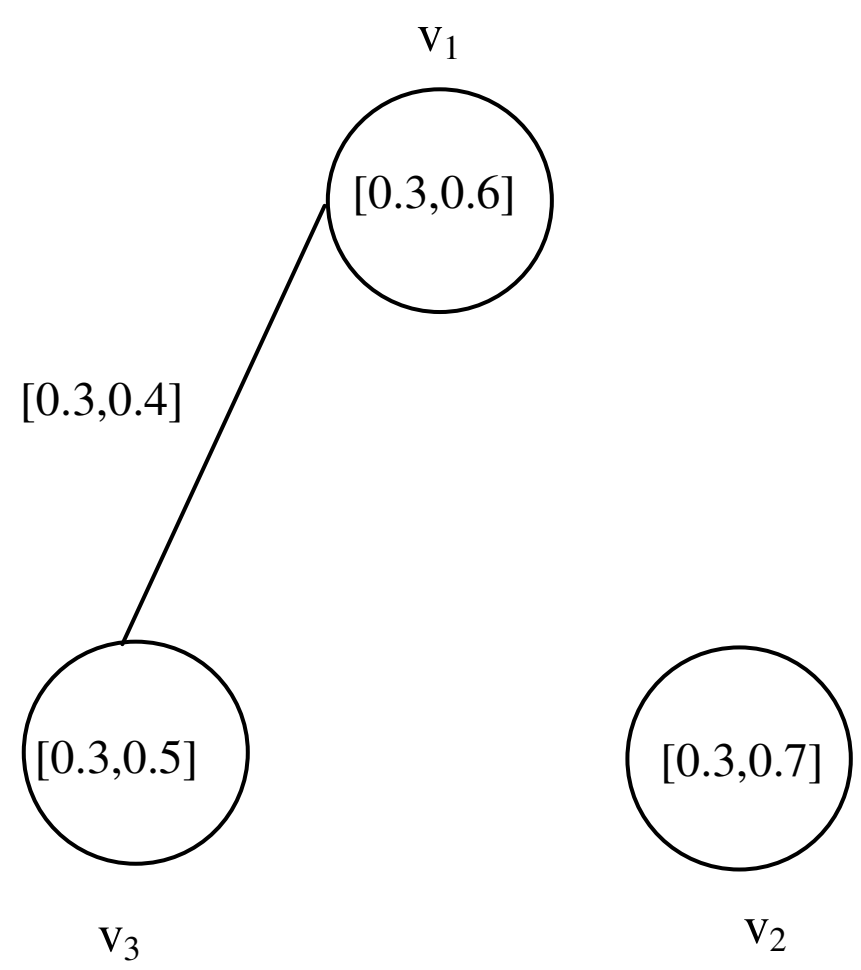

**Fig. 3(b): Antipodal interval - valued fuzzy graph**

By routine calculations, we have:

$\delta_\mu(v_1,v_2)=0.2$ , $\delta_\mu(v_1,v_3)=0.3$ , $\delta_\mu(v_2,v_3)=0.2$ , $\delta_\upsilon(v_1,v_2)=0.9$ , $\delta_\upsilon(v_1,v_3)=1$ , $\delta_\upsilon(v_2,v_3)=0.9$ , $e_\mu(v_1)=0.3$ , $e_\mu(v_2)=0.2$ , $e_\mu(v_3)=0.3$, $e_\upsilon(v_1)=1$, $e_\upsilon(v_2)=0.9$ , $e_\upsilon(v_3)=1$, $d(G)=(0.3,1)=\delta(v_1,v_3)$.

Hence $A(G)=(E,F)$, such that $E=\{v_1,v_2,v_3\}$ and $F=\{v_1v_3\}$.

**Theorem 3.4:** Let $G=(A,B)$ be a complete interval - valued fuzzy graph where $(\mu_A,\upsilon_A)$ is constant function then G is isomorphic to A(G).

**Proof:** Given that $G=(A,B)$ be a complete interval - valued fuzzy graph with $(\mu_A,\upsilon_A)=(k_1,k_2)$, where $k_1$ and $k_2$ are constants, which implies that $\delta(v_i,v_j)=(L_1,L_2)$ , $\forall\ v_i,v_j\in V$. Therefore, eccentricity $e(v_i)=(L_1,L_2)\ \forall\ v_i\in V$, which implies that $d(G)=(L_1,L_2)$. Hence $\delta(v_i,v_j)=(L_1,L_2)=d(G)$ , $\forall\ v_i,v_j\in V$. Hence every pair of vertices are made as neighbors in A(G) such that

(i) An interval - valued fuzzy vertex set of G is taken as interval - valued fuzzy vertex set of A(G), that is, $\mu_F(v_i) = \mu_A(v_i)$ and $\upsilon_E(v_i) = \upsilon_A(v_i)$ for all $v_i \in V$ ,

(ii) $\mu_F(v_i v_j) = \mu_B(v_i v_j)$, since $v_i$ and $v_j$ are neighbors in G and $\upsilon_F(v_i v_j) = \upsilon_B(v_i v_j)$, since $v_i$ and $v_j$ are neighbors in G.

It has same number of vertices, edges and it preserves degrees of the vertices. Hence $G \cong A(G)$.

**Theorem 3.5:** Let $G = (A, B)$ is a connected interval- valued fuzzy graph. Every antipodal interval - valued fuzzy graph is spanning subgraph of G.

**Proof:** By the definition of an antipodal interval - valued fuzzy graph, A(G) contains all the vertices of G. That is ,

(i) $\mu_E(x) = \mu_A(x)$ and $\upsilon_E(x) = \upsilon_A(x)$ for all $x \in V$ and

(ii) If $\delta(x, y) = d(G)$ , then $\mu_F(xy) = \mu_B(xy)$ if x and y are neighbors in G,

$\mu_F(xy) = \min\left(\mu_A(x), \mu_A(y)\right)$ if x and y are not neighbors in G,

$\upsilon_F(xy) = \upsilon_B(xy)$ if x and y are neighbors in G,

$\upsilon_F(xy) = \min\left(\upsilon_A(x), \upsilon_A(y)\right)$ if x and y are not neighbors in G.

Hence A(G) is spanning subgraph of G.

**Definition 3.6:** A homomorphism between two interval- valued fuzzy graphs $G_1 = (A_1, B_1)$ and $G_2 = (A_2, B_2)$ is defined $h: V_1 \to V_2$ is a map which satisfies

(a) $\mu_{A_1}(u) \le \mu_{A_2}(h(u))$ , $\upsilon_{A_1}(u) \le \upsilon_{A_2}(h(u))$ for all $u \in V$

(b) $\mu_{B_1}(uv) \le \mu_{B_2}\left(h(u)\, h(v)\right)$ , $\upsilon_{B_1}(uv) \le \upsilon_{B_2}\left(h(u)\, h(v)\right)$ for all $uv \in E_1$.

**Definition 3.7:** Consider two interval- valued fuzzy graphs $G_1 = (A_1, B_1)$ and $G_2 = (A_2, B_2)$. An isomorphism between two interval - valued fuzzy graphs $G_1$ and $G_2$, denoted by $G_1 \cong G_2$, is a bijective map $h: V_1 \to V_2$ which satisfies

(c) $\mu_{A_1}(u) = \mu_{A_2}(h(u))$ , $\upsilon_{A_1}(u) = \upsilon_{A_2}(h(u))$ for all $u \in V_1$

(d) $\mu_{B_1}(uv) = \mu_{B_2}\left(h(u)\, h(v)\right)$ , $\upsilon_{B_1}(uv) = \upsilon_{B_2}\left(h(u)\, h(v)\right)$ for all $uv \in E_1$.

It is denoted by $G_1 \cong G_2$.

**Definition 3.8:** A co - weak isomorphism between two interval - valued fuzzy graphs $G_1=(A_1,B_1)$ and $G_2=(A_2,B_2)$ is defined as $h:V_1\to V_2$ is a bijective homomorphism that satisfies $\mu_{B_1}(uv)=\mu_{B_2}(h(u)\,h(v))$ and $\upsilon_{B_1}(uv)=\upsilon_{B_2}(h(u)\,h(v))$.

**Definition 3.9:** An interval - valued fuzzy graphs $H=(A',B')$ is said to be an interval - valued fuzzy subgraph of a connected interval - valued fuzzy graph $G=(A,B)$, if $\mu'_A(v_i)=\mu_A(v_i)$ , $\upsilon'_A(v_i)=\upsilon_A(v_i)$ , $\forall\ v_i\in V'$ and $\mu'_B(v_iv_j)=\mu_B(v_iv_j)$ and $\upsilon'_B(v_iv_j)=\upsilon_B(v_iv_j)$ , $\forall\,(v_i,v_j)\in E'$.

**Example 3.10:** In the following figure, we show the interval - valued fuzzy graph $G=(A,B)$ and its subgraph $H=(A',B')$, such that $V=\{v_1,v_2,v_3,v_4\}$, $E=\{(v_1,v_2),(v_2,v_3),(v_3,v_4),(v_4,v_2),(v_4,v_5),(v_5,v_1)\}$, $V'=\{v_2,v_3,v_4\}$, $E'=\{(v_2,v_3),(v_3,v_4),(v_4,v_2)\}$, where $\mu'_A(v_i)=\mu_A(v_i)$ , $\upsilon'_A(v_i)=\upsilon_A(v_i)$ , $\forall\ v_i\in V'$ and $\mu'_B(v_iv_j)=\mu_B(v_iv_j)$ and $\upsilon'_B(v_iv_j)=\upsilon_B(v_iv_j)$, $\forall\ (v_i,v_j)\in E'$.

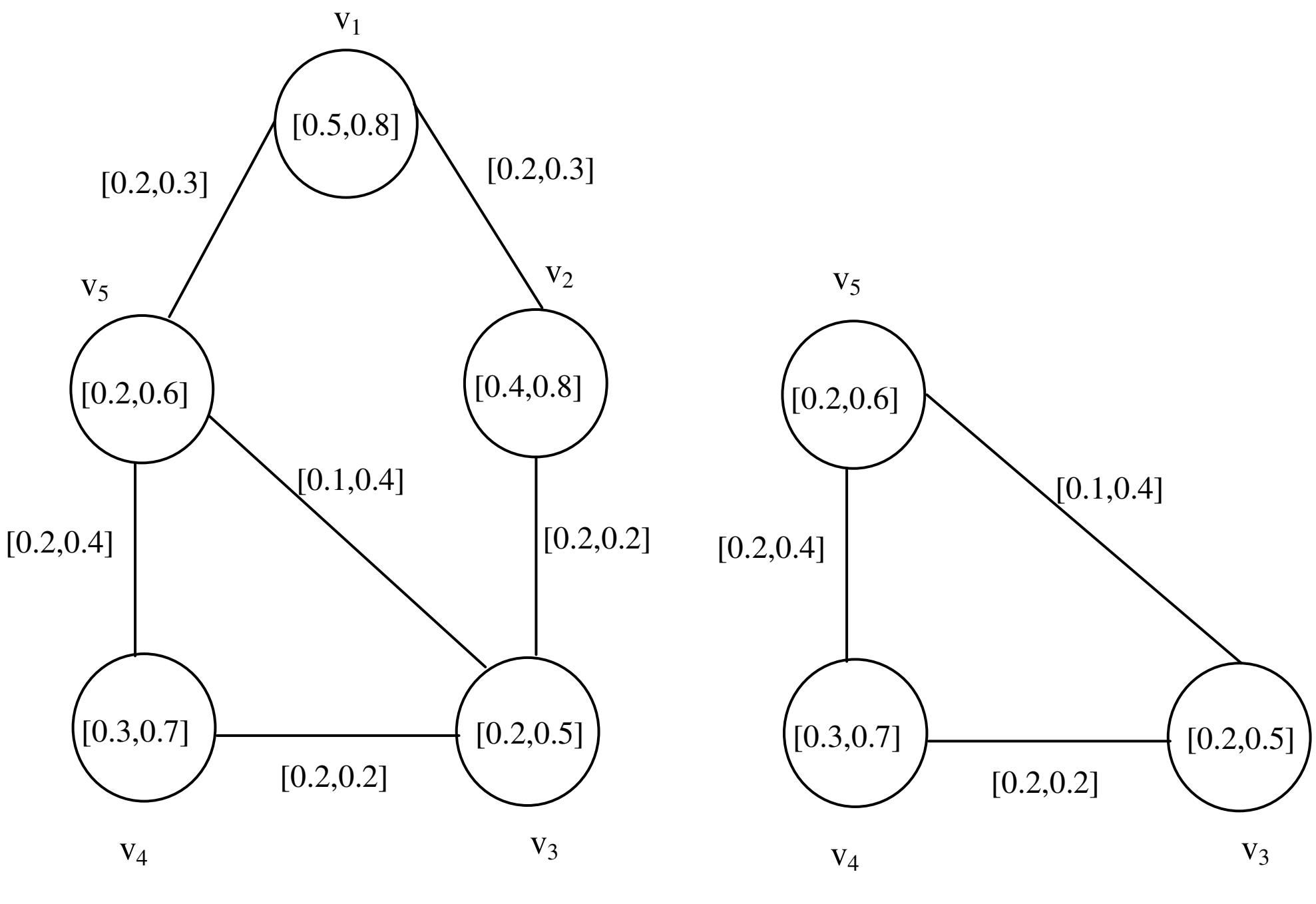

**Fig 4: Interval - valued fuzzy graph and its subgraph**

**Theorem 3.11:** If $G_1$ and $G_2$ are isomorphic to each other, then $A(G_1)$ and $A(G_2)$ are also isomorphic.

**Proof:** As $G_1$ and $G_2$ are isomorphic, the isomorphism h, between them preserves the edge weights, so the $\mu\upsilon$ -length and $\mu\upsilon$ -distance will also be preserved.
Hence if the vertex $v_i$ has the maximum $\mu$ -eccentricity and max $\upsilon$ -eccentricity, in $G_1$, then $h(v_i)$ has the maximum $\mu$ -eccentricity and maximum $\upsilon$ -eccentricity, in $G_2$. So $G_1$ and $G_2$ will have the same diameter.
If the $\mu\upsilon$ -distance between $v_i$ and $v_j$ is $(k_1,k_2)$ in $G_1$ , then $h(v_i)$ and $h(v_j)$ will also have their $\mu\upsilon$ -distance as $(k_1,k_2)$. The same mapping h itself is a bijection between $A(G_1)$ and $A(G_2)$ satisfying the isomorphism condition.

(i) $\mu_{E_1}(v_i)=\mu_{A_1}(v_i)=\mu_{A_2}(h(v_i))=\mu_{E_2}(h(v_i)),\ \forall\, v_i\in G_1$

(ii) $\upsilon_{E_1}(v_i)=\upsilon_{A_1}(v_i)=\upsilon_{A_2}(h(v_i))=\upsilon_{E_2}(h(v_i)),\ \forall\, v_i\in G_1$

(iii) $\mu_{F_1}(v_iv_j)=\mu_{B_1}(v_iv_j)$ , if $v_i$ and $v_j$ are neighbors in $G_1$

$\mu_{F_1}(v_iv_j)=\min\left(\mu_{E_1}(v_i)\ ,\ \mu_{E_1}(v_j)\right)$ , if $v_i$ and $v_j$ are not neighbors in $G_1$

(iv) $\upsilon_{F_1}(v_iv_j)=\upsilon_{B_1}(v_iv_j)$ if $v_i$ and $v_j$ are neighbors in $G_1$

$\upsilon_{F_1}(v_iv_j)=\min\left(\upsilon_{E_1}(v_i)\ ,\ \upsilon_{E_1}(v_j)\right)$, if $v_i$ and $v_j$ are not neighbors in $G_1$.

As $h:G_1\to G_2$ is an isomorphism, $\upsilon_{F_1}(v_iv_j)=\upsilon_{B_2}\left(h(v_i)\ h(v_j)\right)$, if $v_i$ and $v_j$ are neighbors in $G_1$. $\upsilon_{F_1}(v_iv_j)=\min\left(\upsilon_{B_2}(v_i)\ ,\ \upsilon_{B_2}(v_j)\right)$, if $v_i$ and $v_j$ are not neighbors in $G_1$. Hence, $\mu_{F_1}(v_iv_j)=\mu_{F_2}\left(h(v_i)\,h(v_j)\right)$ and $\upsilon_{F_1}(v_iv_j)=\upsilon_{F_2}\left(h(v_i)\,h(v_j)\right)$. So, the same h is an isomorphism between $A(G_1)$ and $A(G_2)$.

**Theorem 3.12:** If $G_1$ and $G_2$ are complete interval - valued fuzzy graph such that $G_1$ is co-weak isomorphic to $G_2$ then $A(G_1)$ is co-weak isomorphic to $A(G_2)$.

**Proof:** As $G_1$ is co-weak isomorphism to $G_2$, there exists a bijection $h: G_1 \to G_2$ satisfying, $\mu_{A_1}(v_i) \le \mu_{A_2}(h(v_i))$ , $\mu_{B_1}(v_i v_j) = \mu_{B_2}(h(v_i)\, h(v_j))$ , $\forall\ v_i, v_j \in V_1$.

If $G_1$ has n vertices, arrange the vertices of $G_1$ in such a way that $\mu_{A_1}(v_1) \le \mu_{A_1}(v_2) \le \mu_{A_1}(v_3) \le ... \le \mu_{A_1}(v_n)$. As $G_1$ and $G_2$ are complete, co-weak isomorphic interval- valued fuzzy graph, $\mu_{B_1}(v_i v_j) = \mu_{B_2}(h(v_i)\, h(v_j))$, $\forall\ v_i, v_j \in V_1$. By Theorem 3.4 and the definition of antipodal interval - valued fuzzy graph, we have $A(G_i)$ contains all the vertices of G, where i=1,2. That is, $\mu_F(x) = \mu_A(x)$ and $\upsilon_E(x) = \upsilon_A(x)$ for all $x \in V$ and $\mu_{F_1}(v_i v_j) = \mu_{F_2}(h(v_i)\, h(v_j))$, $\forall\ v_i, v_j \in V_1$. So, the same bijection h is co-weak isomorphism between $A(G_1)$ and $A(G_2)$.

**Theorem 3.13:** If $G_1$ and $G_2$ are complete interval - valued fuzzy graphs such that $G_1$ is co-weak isomorphic to $G_2$, then $A(G_1)$ is homomorphism to $A(G_2)$.

**Proof:** As $G_1$ is co-weak isomorphic to $G_2$, there exists a bijection $h: G_1 \to G_2$ satisfying, $\mu_{A_1}(v_i) \le \mu_{A_2}(h(v_i))$, $\mu_{B_1}(v_i v_j) = \mu_{B_2}(h(v_i)\, h(v_j))$ , $\forall\ v_i, v_j \in V_1$ and $\upsilon_{A_1}(v_i) \le \upsilon_{A_2}(h(v_i)), \upsilon_{B_1}(v_i v_j) = \upsilon_{B_2}(h(v_i)\, h(v_i))$, $\forall\ v_i, v_j \in V_1$. So the $\mu\upsilon$-distance and hence diameter will be preserved.

Let $d(G_1) = d(G_2) = (k_1, k_2)$. If $v_i, v_j \in V_1$ are at a distance $(k_1, k_2)$ in $G_1$, then they are made as neighbors in $A(G_1)$. So, $h(v_i), h(v_j)$ in $G_2$ are also at a $\mu\upsilon$-distance $(k_1, k_2)$ in $G_2$ and $h(v_i), h(v_j)$ are made as neighbors in $A(G_2)$. If $v_i$ and $v_j$ are neighbors then

$$\mu_{F_1}(v_i v_j) = \mu_{B_1}(v_i v_j) = \mu_{B_2}(h(v_i)\, h(v_j)) = \mu_{F_2}(h(v_i)\ \ h(v_j)) ,$$

$$\upsilon_{F_1}(v_i v_j) = \upsilon_{B_1}(v_i v_j) = \upsilon_{B_2}(h(v_i)\, h(v_j)) = \upsilon_{F_2}(h(v_i)\, h(v_j)).$$

If $v_i$ and $v_j$ are not neighbors in $G_1$, then

$$\mu_{F_1}(v_i v_j) = \min(\mu_{A_1}(v_i), \mu_{A_1}(v_j)) \le \min(\mu_{A_2}(h(v_i), \mu_{A_2}(h(v_j))) = \mu_{F_2}(h(v_i)\, h(v_j)),$$

$\upsilon_{F_1}(v_i v_j) = \min\left(\upsilon_{A_1}(v_i), \upsilon_{A_1}(v_j)\right) \leq \min\left(\upsilon_{A_2}(h(v_i)), \upsilon_{A_2}(h(v_j))\right) = \upsilon_{F_2}\left(h(v_i)\, h(v_j)\right)$

Hence $A(G_1)$ is homoingmorphism to $A(G_2)$.

**Definition 3.14:** Let $G$ be a connected interval - valued fuzzy graph. The $\mu$-status of a vertex $v_i$ is denoted by $S_\mu(v_i)$ and is defined as $S_\mu(v_i) = \sum_{v_j \in V} \delta_\mu(v_i, v_j)$.

**Definition 3.15:** Let G be a connected interval - valued fuzzy graph. The $\upsilon$-status of a vertex $v_i$ is denoted by $S_\upsilon(v_i)$ and is defined as $S_\upsilon(v_i) = \sum_{v_j \in V} \delta_\upsilon(v_i, v_j)$.

**Definition 3.16:** Let G be a connected interval - valued fuzzy graph. The $\mu\upsilon$-status of a vertex $v_i$ is denoted by $S_{\mu_\upsilon}(v_i)$ and is defined as $S_{\mu_\upsilon}(v_i) = \left(S_\mu(v_i), S_\upsilon(v_i)\right)$.

**Example 3.17:** Consider the following interval - valued fuzzy graph $G = (A, B)$:

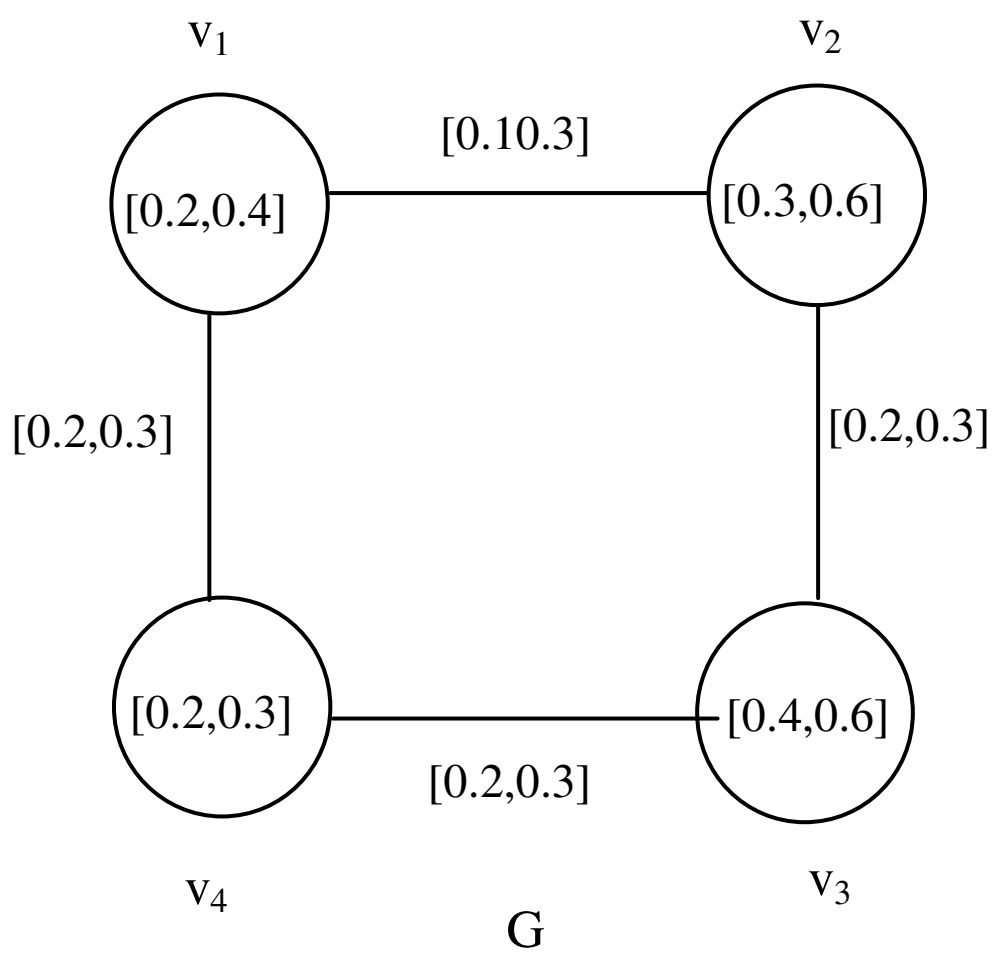

**Fig. 5: Interval - valued fuzzy graph G**

$\delta_\mu(v_1, v_2) = 0.1$, $\delta_\mu(v_1, v_3) = 0.3$, $\delta_\mu(v_1, v_4) = 0.2$, $\delta_\mu(v_2, v_3) = 0.2$,

$\delta_\mu(v_2, v_4) = 0.3$, $\delta_\mu(v_3, v_4) = 0.2$, $\delta_\upsilon(v_1, v_2) = 0.9$, $\delta_\upsilon(v_1, v_3) = 0.6$,

$\delta_\upsilon(v_1, v_4) = 0.9$, $\delta_\upsilon(v_2, v_3) = 0.9$, $\delta_\upsilon(v_2, v_4) = 0.6$, $\delta_\upsilon(v_3, v_4) = 0.9$.

$\delta_{\mu}(v_1) = 0.5$ , $\delta_{\mu}(v_2) = 0.6$ , $\delta_{\mu}(v_3) = 0.7$ , $\delta_{\mu}(v_4) = 0.7$ ,

$\delta_{\upsilon}(v_1) = 2.4$ , $\delta_{\upsilon}(v_2) = 2.4$ , $\delta_{\upsilon}(v_3) = 2.4$ , $\delta_{\upsilon}(v_4) = 2.4$ .

Therefore, $S_{\mu_{\upsilon}}(v_1) = (0.5 , 2.4)$ , $S_{\mu_{\upsilon}}(v_2) = (0.6 , 2.4)$ , $S_{\mu_{\upsilon}}(v_3) = (0.7 , 2.4)$ ,

$S_{\mu_{\upsilon}}(v_4) = (0.7 , 2.4)$ .

**Definition 3.18:** Let G be a connected interval- valued fuzzy graph. The minimum $\mu$-status of G is denoted by $m[S_{\mu}(G)]$ and is defined as $m[S_{\mu}(G)] = \min \left(S_{\mu}(v_i) , \forall v_i \in V\right)$

**Definition 3.19:** Let G be a connected interval - valued fuzzy graph. The minimum $\upsilon$-status of G is denoted by $m[S_{\upsilon}(G)]$ and is defined as $m[S_{\upsilon}(G)] = \min \left(S_{\upsilon}(v_i) , \forall v_i \in V\right)$.

**Definition 3.20.** Let G be a connected interval - valued fuzzy graph. The minimum $\mu\upsilon$-status of G is denoted by $m[S_{\mu_{\upsilon}}(G)]$ and is defined as

$m[S_{\mu_{\upsilon}}(G)] = \left(m[S_{\mu}(G)] \ , \ m[S_{\upsilon}(G)]\right)$.

**Example 3.21:** Consider the following interval - valued fuzzy graph $G = (A, B)$ :

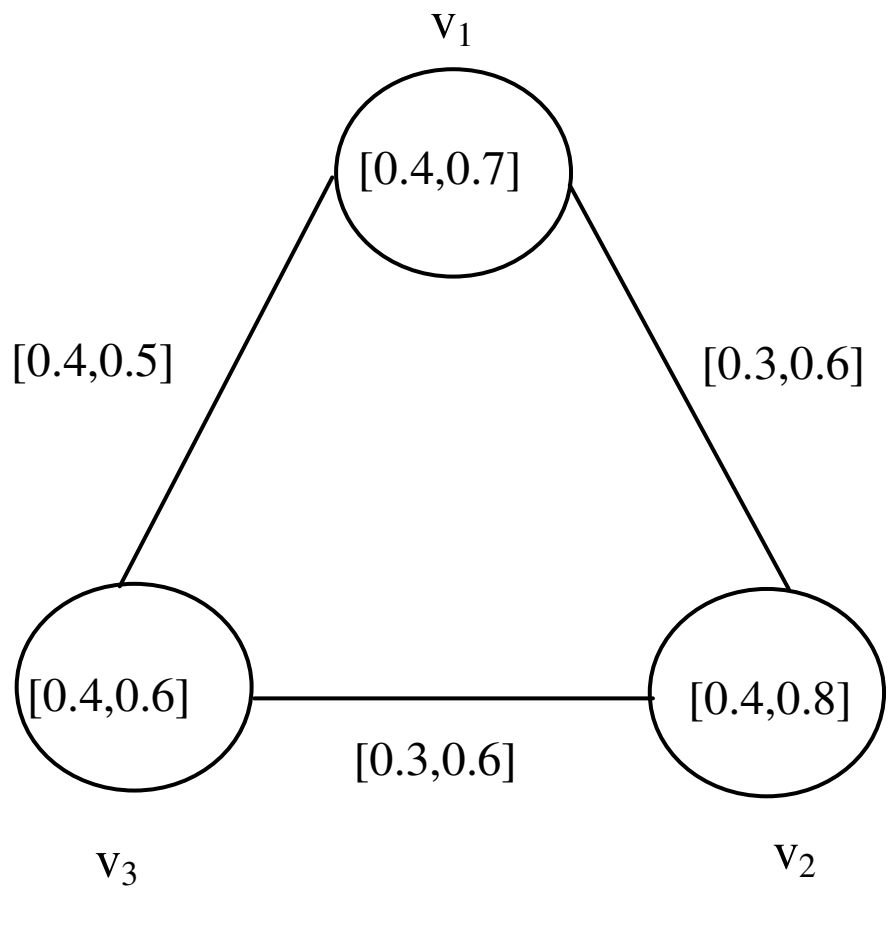

**Fig. 6: Interval - valued fuzzy graph G**

By routine calculations, we have $\delta_{\mu}(v_1, v_2) = 0.3$ , $\delta_{\mu}(v_1, v_3) = 0.4$ , $\delta_{\mu}(v_2, v_3) = 0.3$ ,

$\delta_{\upsilon}(v_1, v_2) = 1.1$ , $\delta_{\upsilon}(v_1, v_3) = 1.2$ , $\delta_{\upsilon}(v_2, v_3) = 1.1$ . $S_{\mu}(v_1) = 0.7$ , $S_{\mu}(v_2) = 0.6$ ,

$S_{\mu}(v_3) = 0.7$ , $S_{\upsilon}(v_1) = 2.3$ , $S_{\upsilon}(v_2) = 2.2$ , $S_{\upsilon}(v_3) = 2.3$ .

Therefore, $S_{\mu_\upsilon}(v_1) = (0.7, 2.3)$ , $S_{\mu_\upsilon}(v_2) = (0.6, 2.2)$ , $S_{\mu_\upsilon}(v_3) = (0.7, 2.3)$ .

$m[S_{\mu_\upsilon}(G)] = (0.6, 2.2)$ .

**Definition 3.22:** Let G be a connected interval - valued fuzzy graph. The maximum $\mu$-status of G is denoted by $M[S_{\mu}(G)]$ and is defined as

$M[S_{\mu}(G)] = \max\left(S_{\mu}(v_i),\ \forall\, v_i \in V\right)$.

**Definition 3.23:** Let G be a connected interval - valued fuzzy graph. The maximum $\upsilon$-status of G is denoted by $M[S_{\upsilon}(G)]$ and is defined as

$M[S_{\upsilon}(G)] = \max\left(S_{\upsilon}(v_i),\ \forall\, v_i \in V\right)$.

**Definition 3.24:** Let G be a connected interval - valued fuzzy graph. The maximum $\mu\upsilon$ status of G is denoted by $M[S_{\mu_\upsilon}(G)]$ and is defined as

$M[S_{\mu_\upsilon}(G)] = \left(M[S_{\mu}(G)], M[S_{\upsilon}(G)]\right)$.

**Example 3.25:** Consider the following interval - valued fuzzy graph $G = (A, B)$ :

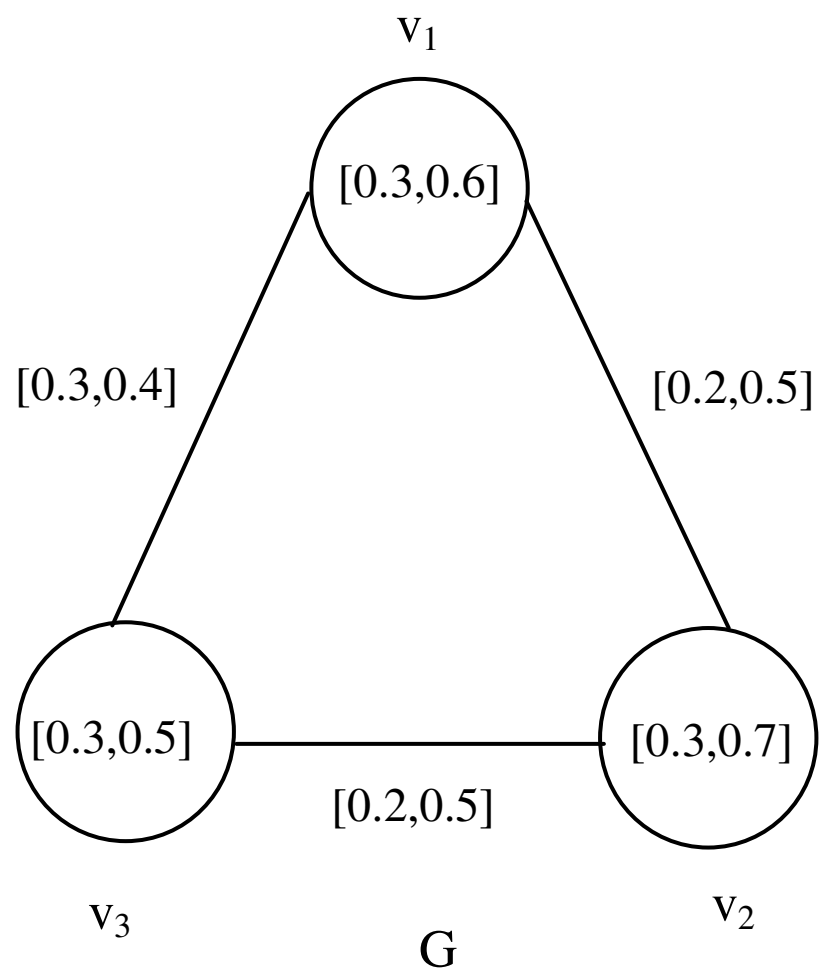

**Fig. 7: Interval - valued fuzzy graph G**

By routine calculations , we have

$\delta_\mu = (v_1, v_2) = 0.2$, $\delta_\mu = (v_1, v_3) = 0.3$, $\delta_\mu = (v_2, v_3) = 0.2$,

$\delta_\upsilon = (v_1, v_2) = 0.9$, $\delta_\upsilon = (v_1, v_3) = 1$, $\delta_\upsilon = (v_2, v_3) = 0.9$, $S_\mu(v_1) = 0.5$,

$S_\mu(v_2) = 0.4$, $S_\mu(v_3) = 0.5$, $S_\upsilon(v_1) = 1.9$, $S_\upsilon(v_2) = 1.8$, $S_\upsilon(v_3) = 1.9$.

Therefore, $S_{\mu_\upsilon}(v_1) = (0.5, 1.9)$, $S_{\mu_\upsilon}(v_2) = (0.4, 1.8)$, $S_{\mu_\upsilon}(v_3) = (0.5, 1.9)$.

$M\,[S_{\mu_\upsilon}(G)] = (0.5, 1.9)$.

**Definition 3.26:** The total $\mu$-status of an interval - valued fuzzy graph G is denoted by $t\,[S_\mu(G)]$ and is defined as $t\,[S_\mu(G)] = \sum_{v_i \in V} S_\mu(v_i)$.

**Definition 3.27:** The total $\upsilon$-status of an interval - valued fuzzy graph G is denoted by $t\,[S_\upsilon(G)]$ defined as $t\,[S_\upsilon(G)] = \sum_{v_i \in V} S_\upsilon(v_i)$.

**Definition 3.28:** The total $\mu\upsilon$-status of an interval - valued fuzzy graph G is denoted by $t\,[S_{\mu_\upsilon}(G)]$ and is defined as $t\,[S_{\mu_\upsilon}(G)] = \left(t\,[S_\mu(G)], t\,[S_\upsilon(G)]\right)$.

**Definition 3.29:** The median of an interval - valued fuzzy graph G is denoted by M(G) and is defined as the set of nodes with minimum $\mu\upsilon$ status.

**Definition 3.30:** An interval - valued fuzzy graph G is said to be self - median if all the vertices have the same status. In order words, G is self - median if and only if $m[S_{\mu_\upsilon}(G)] = M[S_{\mu_\upsilon}(G)]$.

**Example 3.31:** Let $G = (A, B)$ be an interval - valued fuzzy graph defined as follows:

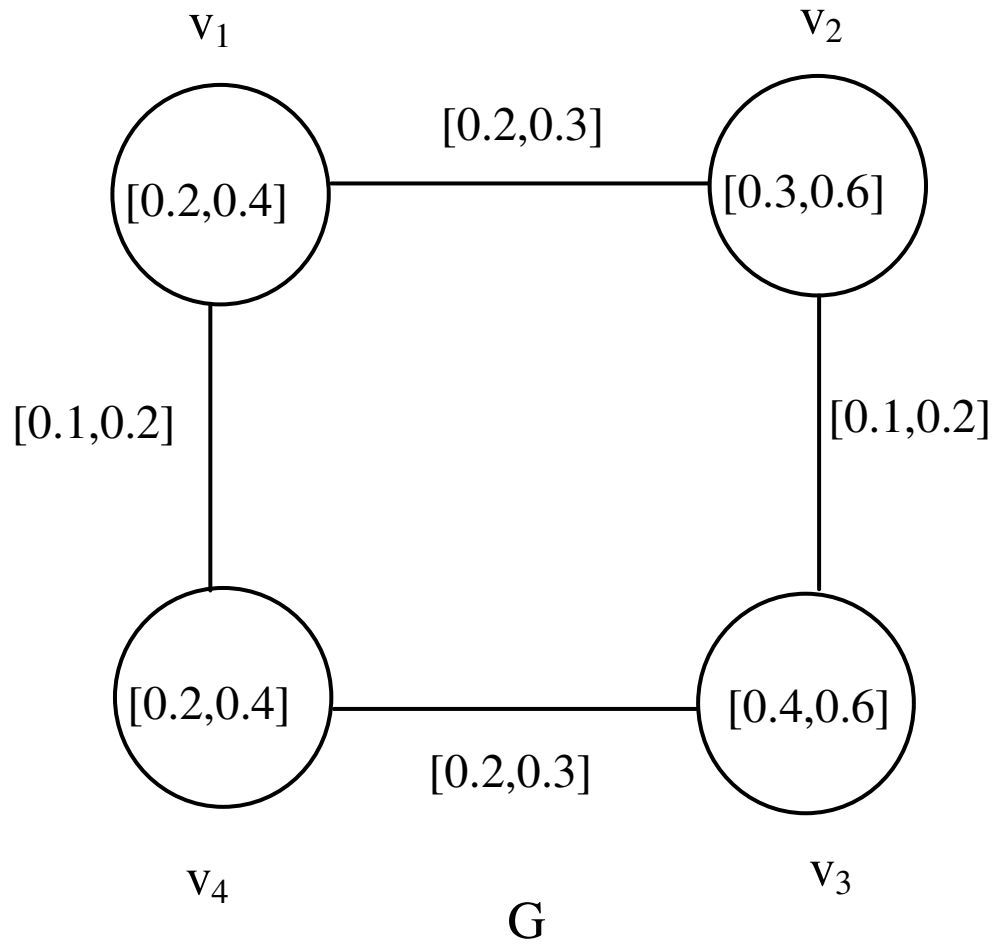

**Fig. 8: Self -median interval - valued fuzzy graph G**

By routine calculations, we have

$\delta_\mu(v_1,v_2)=0.2$ , $\delta_\mu(v_1,v_4)=0.1$, $\delta_\mu(v_1,v_3)=0.3$ , $\delta_\mu(v_2,v_3)=0.1$,

$\delta_\mu(v_2,v_4)=0.3$, $\delta_\mu(v_3,v_4)=0.2$, $\delta_\upsilon(v_1,v_2)=0.7$ , $\delta_\upsilon(v_1,v_3)=0.5$,

$\delta_\upsilon(v_1,v_4)=0.8$, $\delta_\upsilon(v_2,v_3)=0.8$ , $\delta_\upsilon(v_2,v_4)=0.5$, $\delta_\upsilon=(v_3,v_4)=0.7$.

$S_\mu(v_1)=0.6$, $S_\mu(v_2)=0.6$, $S_\mu(v_3)=0.6$, $S_\mu(v_4)=0.6$,

$S_\upsilon(v_1)=2$ , $S_\upsilon(v_2)=2$, $S_\upsilon(v_3)=2$, $S_\upsilon(v_4)=2$. Therefore,

$S_{\mu_\upsilon}(v_1)=(0.6\,,2)$ , $S_{\mu_\upsilon}(v_2)=(0.6\,,2)$ ,

$S_{\mu_\upsilon}(v_3)=(0.6\,,2)$, $S_{\mu_\upsilon}(v_4)=(0.6\,,2)$ and $t[S_{\mu_\upsilon}(G)]=(2.4\,,8)$.

Here, $S_{\mu_\upsilon}(v_i)=(0.6\,,2)$, $\forall\, v_i\in V$. Hence G is self median interval - valued fuzzy graph.

**Theorem 3.32:** Let G be an interval - valued fuzzy graph, where crisp graph $G^*$ is an even cycle. If alternate edges have same membership values and non-membership values, then G is self median interval – valued fuzzy graph.

**Proof.** Given that G is an interval - valued fuzzy graph. Since crisp graph $G^*$ is an even cycle. Also, alternate edges of G have same membership values and non-memebership values, we have

$\delta(v_1,v_2)=\delta(v_3,v_4)=...=\delta(v_{n-1},v_n)$ and similarly,

$\delta(v_2,v_3)=\delta(v_4,v_5)=...=\delta(v_n,v_1)$ , $\delta(v_1,v_3)=\delta(v_2,v_4)=\delta(v_3,v_5)=...=L$.

Hence, $S_\mu(v_i)=k$ and $S_\upsilon(v_i)=m$ , $\forall\ v_i\in V$. So, G is a self median interval - valued fuzzy graph.

**Remark 3.33:** Let G be an interval - valued fuzzy graph, where crisp graph $G^*$ is an odd cycle. If alternate edges have same membership values and non-membership values, then G may not be self median interval - valued fuzzy graph.

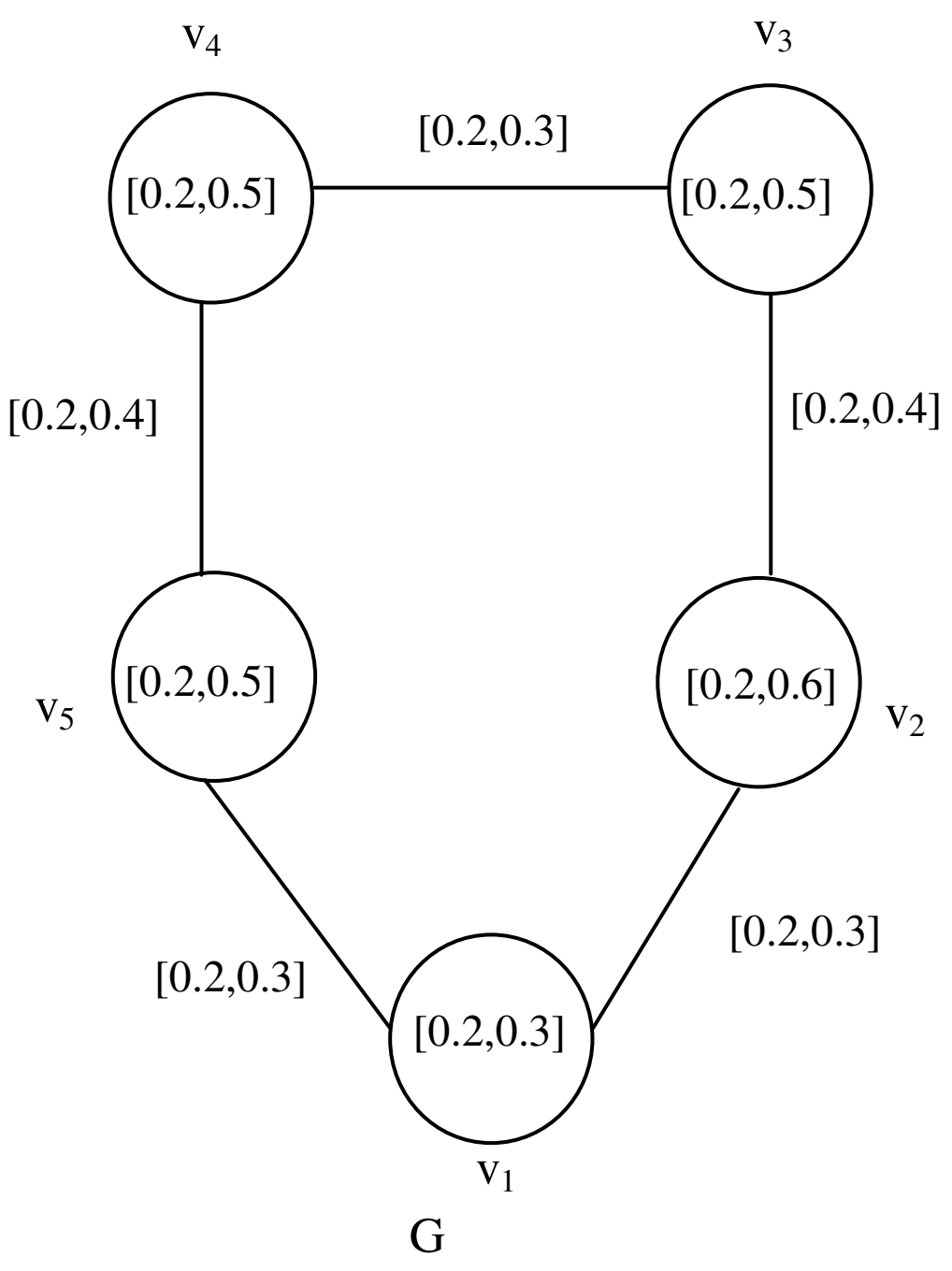

**Fig. 9: Interval - valued fuzzy graph G**

By routine calculations, we have

$\delta_{\mu}(v_1,v_2)=0.2$, $\delta_{\mu}(v_1,v_4)=0.4$, $\delta_{\mu}(v_1,v_5)=0.2$, $\delta_{\mu}(v_2,v_3)=0.2$ ,

$\delta_{\mu}(v_1,v_3)=0.7$, $\delta_{\mu}(v_2,v_4)=0.4$, $\delta_{\mu}(v_2,v_5)=0.4$ , $\delta_{\mu}(v_3,v_4)=0.2$ ,

$\delta_{\mu}(v_3,v_5)=0.4$, $\delta_{\mu}(v_4,v_5)=0.2$, $\delta_{\upsilon}(v_1,v_2)=1.4$ , $\delta_{\upsilon}(v_1,v_3)=1$ ,

$\delta_{\upsilon}(v_1,v_4)=1$, $\delta_{\upsilon}(v_1,v_5)=1.4$, $\delta_{\upsilon}(v_2,v_3)=1.3$, $\delta_{\upsilon}(v_2,v_4)=1$, $\delta_{\upsilon}(v_2,v_5)=1.1$,

$\delta_{\upsilon}(v_3,v_4)=1.4$, $\delta_{\upsilon}(v_3,v_5)=1$, $\delta_{\upsilon}(v_4,v_5)=1.3$ .

That is , $\delta(v_1,v_2)=(0.2,1.4)$, $\delta(v_1,v_3)=(0.7,1)$, $\delta(v_1,v_4)=(0.4,1)$ ,

$\delta(v_1,v_5)=(0.2,1.4)$ , $\delta(v_2,v_3)=(0.2,1.3)$

, $\delta(v_2,v_4)=(0.4,1)$, $\delta(v_2,v_5)=(0.4,1.1)$ , $\delta(v_3,v_4)=(0.2,1.4)$

, $\delta(v_3,v_5)=(0.4,1)$, $\delta(v_4,v_5)=(0.2,1.3)$. $S_{\mu_{\upsilon}}(v_1)=(1.5,4.8)$,

$S_{\mu_{\upsilon}}(v_2)=(1.2,4.8)$, $S_{\mu_{\upsilon}}(v_3)=(1.5,4.7)$, $S_{\mu_{\upsilon}}(v_4)=(1.2,4.7)$, $S_{\mu_{\upsilon}}(v_5)=(1.2,4.8)$.

Here, $S_{\mu_\upsilon}(v_1) \neq S_{\mu_\upsilon}(v_3)$ , $S_{\mu_\upsilon}(v_1) \neq S_{\mu_\upsilon}(v_4)$ and $S_{\mu_\upsilon}(v_2) \neq S_{\mu_\upsilon}(v_3)$ , $S_{\mu_\upsilon}(v_2) \neq S_{\mu_\upsilon}(v_4)$ .

That is, all the vertices does not have the same status. Hence G is not a self median interval - valued fuzzy graph.

## 4. Conclusions

It is known that fuzzy graph theory has numerous applications in modern science and engineering, especially in the field of information theory, neural networks, expert systems, cluster analysis, medical diagnosis, traffic engineering, network routing, town planning, and control theory. In this paper, we have introduced the concept of antipodal interval- valued fuzzy graphs and self median interval - valued fuzzy graphs of the given interval - valued fuzzy graphs. We investigated isomorphism properties on antipodal interval-valued fuzzy graphs. In our future work, we will focus on direct sum of two interval - valued fuzzy graphs and will study the truncations of the direct sum of two interval - valued fuzzy graphs.

**Acknowledgement:** The authors are grateful to the reviewers for the suggestions to improvement of the presentation of the paper.